\documentstyle{mn}

\input epsf

\def\spose#1{\hbox to 0pt{#1\hss}}
\newcommand{\approxlt}{\mathrel{\spose{\lower 3pt\hbox{$\sim$}}
	\raise 2.0pt\hbox{$<$}}}
\newcommand{\approxgt}{\mathrel{\spose{\lower 3pt\hbox{$\sim$}}
	\raise 2.0pt\hbox{$>$}}}
\newcommand{\ergpcmsqps}{{\rm erg}\,{\rm cm}^{-2}\,{\rm s}^{-1}\,}

\title[Soft versus Hard X--Ray Emission in AGN]
{Soft versus Hard X--Ray Emission in AGN: Partial Covering and Warm plus Cold Absorber Models}

\author[M.T. Ceballos \& X. Barcons]
	{ M.T. Ceballos$^{1,2}$ \& X. Barcons$^{1}$
	\\
	$^1$ Instituto de F\'{\i}sica de Cantabria (Consejo Superior
	de Investigaciones Cient\'{\i}ficas - Universidad\\ 
	de Cantabria), 39005 Santander, Spain\\
	$^2$ Departamento de F\'{\i}sica Moderna, Universidad de
	Cantabria, 39005 Santander, Spain}

\begin{document}

\maketitle

\begin{abstract}      
 
We analyse the {\it ROSAT} PSPC hardness ratio and the 0.5-2 keV to
2-10 keV flux ratio of 65 Active Galactic Nuclei (AGN) for which there
are both {\it ROSAT} archival observations available and 2-10 keV
fluxes, mostly from the HEAO-1 MC-LASS survey.  We conclude that the
simplest spectral model for the AGN that can accommodate the variety
of X-ray colours obtained is a standard power law (with energy
spectral index $\alpha\sim 0.9$) plus a $\sim 0.1$ keV black body both
partially absorbed.  In our sample, type 1 AGN require an absorbing
column around $10^{22}\, {\rm cm}^{-2}$ with covering fractions
between 20 and 100\%, while type 2 AGN display larger columns and
$\sim 100\%$ coverage. This simple model also provides a good link
between soft and hard AGN X-ray luminosity functions and source
counts. We also consider a warm absorber as an alternative model to
partial covering and find that the the presence of gas in two phases
(ionized and neutral) is required.

\end{abstract}


\section{INTRODUCTION}

Comparing the information obtained through the analyses of AGN data in
different X-ray energy bands appears to be a difficult task which
leads in some cases to striking results. For example, 2-10 keV X-ray
source counts (which are dominated by AGN) obtained by the {\it Ginga}
fluctuation analyses (Warwick \& Butcher 1992) are clearly above the
number counts obtained from the {\it Einstein Observatory Medium
Sensitivity Survey} (EMSS) in the 0.3-3.5 keV band (Gioia et al. 1990)
if a power law spectrum with energy spectral index $\alpha\approxgt
0.7$ and negligible photoelectric absorption are assumed.  Several
explanations have been proposed to bring these results into
consistency. Warwick \& Butcher (1992) were able to fit the spectrum
of the fluctuations in the 2-10 keV band (which should be close to the
median X-ray spectrum of a source at the level where there is one
source per {\it Ginga} LAC beam $\sim 5\times 10^{-13}\, \ergpcmsqps$)
to a power law with an energy spectral index $\alpha \approx 0.8$ with
no evidence for photoelectric absorption (the derived upper limit is
$N_{H}\approxlt 3 \times 10^{21} \rm cm^{-2}$, Stewart 1992). In
order to have a 2-10 keV to 0.3-3.5 keV flux ratio of about 2 (which
is what is needed in order to reconcile source counts in both energy
bands) a photoelectric absorption larger than the above upper limit is
required. In this case, the contribution of clusters of galaxies,
having a much softer spectrum will compensate the AGN photoelectric
absorption in the spectrum of the fluctuations (Barcons 1993). If
photoelectric absorption were ignored, an energy spectral index
$\alpha \sim 0.4$, much flatter than the typical index for any class
of source (and this indeed includes AGN) at that flux level, would be
required. Therefore, absorption in AGN appears to be a necessity to
solve the soft/hard X-ray source counts discrepancies.

On the other hand, soft X-ray selected AGN actually exhibit low-energy
excesses over the average power law (Maccacaro et al. 1988, Turner and
Pounds 1989, Hasinger 1992).  Franceschini et al. (1993) proposed a
scenario to account for these facts: the existence of two different
populations, the soft X--ray sources, with steep spectrum and high
evolution rates and the hard X-ray sources, with a weak cosmological
evolution and strong self-absorption. Recent models for the origin of
the X-ray background (Madau, Ghisellini \& Fabian 1994, Comastri et al
1995) based on the AGN unified scheme (Antonucci \& Miller 1985,
Antonucci 1993), rather suggest that there is a continuity between
these two populations.

We analyse a sample of AGN for which there are 2-10 keV fluxes (mostly
coming from the HEAO-1 MC-LASS survey, Wood et al 1984) and archival
{\it Rosat} PSPC observations. By analysing the PSPC hardness ratios
versus 0.5-2 keV to 2-10 keV flux ratio, we conclude that the simplest
spectral model that can accommodate the whole sample is a power law
plus a blackbody both partially absorbed. 

Furthermore the model we propose is able to bring into consistency the
source counts as well as the AGN luminosity functions in both
bands. Our comparison is relevant only to local ($z<0.2$) AGN and its
extension to higher redshifts would require more data.

We also consider an alternative to this partial covering scenario
which can account for the spectra of AGN as well as for the
soft excess observed: a full obscuration of the X--ray continuum by
partly ionized gas (see Netzer 1993 and references therein). However
unless some neutral absorbing material is also present the spectra are
invariably too soft.

In section 2 we describe the sample, present the broad-band hardness
ratio versus flux ratio relation, and introduce the simplest model
also able to accommodate the wide range of parameter space occupied by
these sources. The warm absorber model is also introduced in this
section as an alternative model to describe the spread in the observed
parameters. In section 3 we show that a partial covering model is able
to bring soft and hard X-ray AGN luminosity functions and source
counts into agreement. We summarize the results and present some
conclusions in section 4.

\section{ BROAD BAND HARDNESS RATIOS}

\subsection{The sample}

In order to study the Broad Band X-ray colours, we constructed a local
(z$<$0.2) AGN sample, in such a way that $K$-corrections or
evolutionary effects do not come into play.  We tried to build the
largest sample for which there is a 2-10 keV flux measurement and {\it
Rosat} PSPC information contained in the WGACAT point source catalogue
in the public archive at {\it HEASARC}. The resulting sample is listed
in Table 1.

Most of the objects come from the LMA sample described by Grossan
(1992), which is the AGN sample of the HEAO-1 MC-LASS, from which we
take the 2-10 keV fluxes. Other AGN with 2-10 keV fluxes measured
either by {\it Ginga} (19 sources, Awaki 1992, Nandra and Pounds 1994,
Turner et al 1992b) or {\it EXOSAT} (5 sources, Turner \& Pounds 1989)
have also been included. However our conclusions remain unaffected if
these low luminosity objects (the {\it Ginga} and {\it EXOSAT} ones)
are removed. From the LMA sample we had to eliminate two sources known
to be contaminated by another nearby source: III Zw 2 (Tagliaferri et
al 1988) and 3A0057-383 (Giommi et al 1989, George et al 1995). From
the EXOSAT sub-sample we also eliminate 3C445 which is close to the
cluster of galaxies A2440 which can contaminate its 2--10 keV
flux. The WGACAT point source catalogue provides the counts collected
by the PSPC in the soft (11-39) PI channels and in the hard (40-200)
PI channels. The 0.5-2 keV fluxes have been obtained from these data,
under the assumption of an energy spectral index of $\alpha=0.9$ and
Galactic photoelectric absorption (also given in the WGACAT point
source catalogue), although our conclusions remain unaffected if other
model spectra are assumed.

Indeed, the 2-10 keV and {\it Rosat} PSPC observations are not
simultaneous and therefore variability could be relevant for each
source individually. However, we do not intend here to derive specific
spectral properties for each source, but we rather regard the AGN
sample as an ensemble and therefore no systematic variations in the
X-ray colours are expected from the variability. But even for
individual sources, a factor of 2 variation in the flux ratios (see
below) will not affect our conclusions.

\subsection{Hardness ratios.}

We defined a source hardness ratio in the soft band as:
\begin{equation}
HR=\frac{H-S}{H+S}
\end{equation}
$S$ being the number of counts in ROSAT PI channels 11-39
(0.1-0.4 keV) and $H$ the number of counts in channels 40-200 (0.4-2.0
keV).

In Figure 1 we plot the PSPC hardness ratio (corrected for Galactic
absorption) versus flux ratio, defined as the ratio between the soft
(0.5-2 keV) band flux and the hard band flux (2-10 keV), for the
sample described above. The Galactic correction is strongly dependent
on the model assumed so we used the spectral model described below in
order to be self-consistent. Maybe the most striking feature of this
plot is the large spread of the AGN population. That indeed implies
that there is no universal AGN spectrum.

\begin{figure*}
  \vbox to 0cm{\vfil}
  \label{fig1}
\epsfverbosetrue
\epsfysize= 300pt
\epsffile{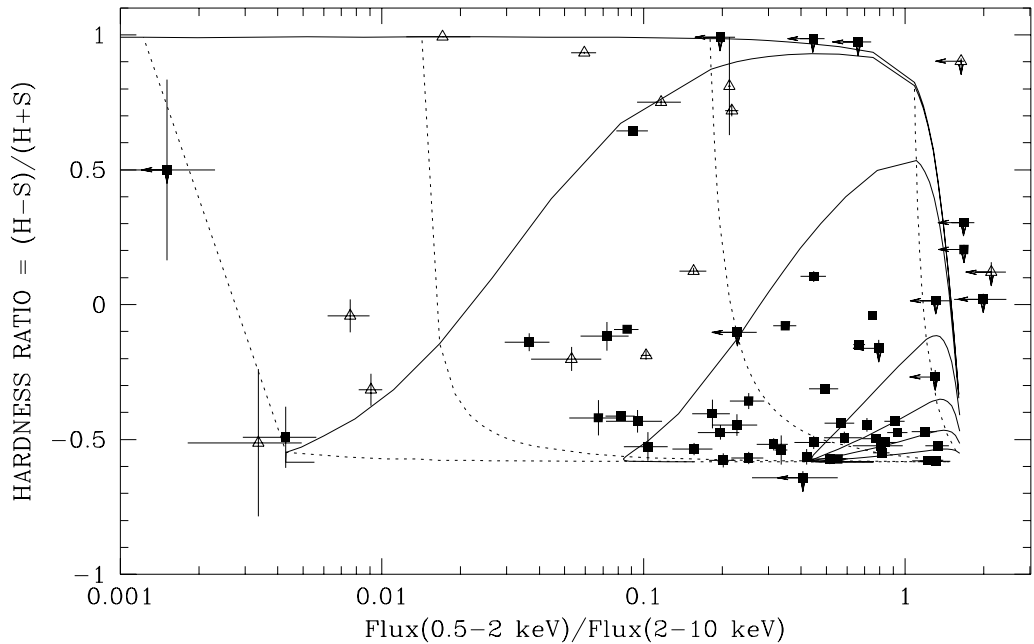}
\caption{ROSAT PSPC hardness ratio (corrected for Galactic absorption
according to the spectral model explained in 2.2) versus
flux ratio.  $H$ and $S$ are the counts in the PI channels 40-200 and
 11-39 respectively.  The individual points represent the sources in
 the sample, Seyfert 1 or QSO (filled squares) and Seyfert 2 (open
 triangles). Along each solid line the covering factor in the
 simulation is kept constant whereas the column density parameter
 changes from $10^{20} \rm cm^{-2}$ to $3\times 10^{23} \rm cm^{-2}$
 from right to left. The values assumed for the covering factor are
 $f_{cov}$= 0.20 (lower solid line), 0.40, 0.60, 0.80, 0.97, 0.999, 1.00
 (uppermost curve). The dotted lines have constant column density
 values, $N_{H}= 10^{21}$ (dotted line on the right), $10^{22}$, $
 5\times 10^{22}$, $10^{23}$ (dotted line on the left) and the
 covering factor ranges from 0.2 to 1.0 from bottom to top along the
 line.}
\end{figure*}

In order to explain this colour-colour diagram exhibited by the
sources we carried out some simulations with XSPEC, assuming model
spectra for the sources and folding them through the ROSAT PSPC
response matrix to obtain the counts in each channel (from
which we compute $S$ and $H$) as well as the fluxes in the soft (0.5-2
keV) and hard (2-10 keV) energy bands.  

The simplest model for the AGN spectrum, an unabsorbed power law,
turns out to be unable to reproduce the scatter observed in the
hardness ratios even if the energy spectral index is allowed to vary.
This is clearly shown in figure 2 where we show the expected position
in the X-ray colour diagram by this model for a couple of typical
energy spectral indices.

\begin{figure*}
  \vbox to 0cm{\vfil}
  \label{fig2}
\epsfverbosetrue
\epsfysize= 300pt
\epsffile{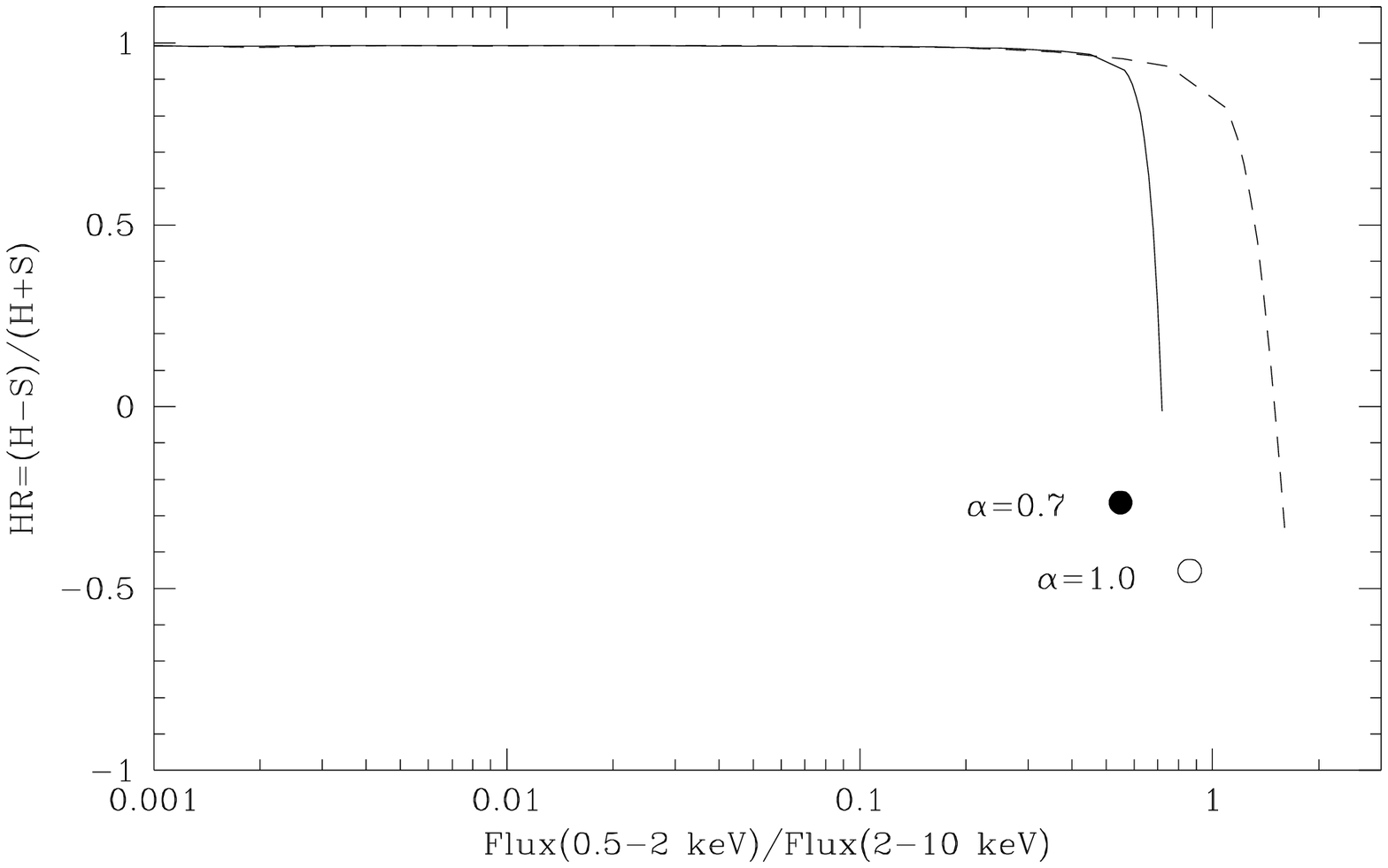}
  \caption{Simulated hardness ratio versus flux ratio for different
spectral models. Filled circle: power law with $\alpha=0.7$. Hollow
circle: power law with $\alpha=1.0$. Solid line: absorbed power law with
$\alpha=0.9$ and column density ranging from $10^{20}$ to
$10^{24}\,\rm cm^{-2}$ from right to left along the line. Dashed line:
absorbed power law with $\alpha=0.9$, blackbody temperature $kT=0.1$
keV and contribution of 50\% of the power law at 1 keV and column
density ranging from $10^{20}$ to $10^{24}\,\rm cm^{-2}$ from right to
left along the line.}
\end{figure*}

The next step, is to assume some intrinsic photoelectric absorption
for the AGN.  The solid line in Figure 2 is obtained by varying the
column density for an $\alpha=0.9$ standard power law, from the
Galactic value to $\sim 10^{24}\, {\rm cm}^{-2}$. With this model,
a wider range of hardness ratios would be covered. Moreover higher
flux ratios ( F(0.5--2)/F(2--10)$\ge$0.6) could be obtained if steeper
power laws were taken into account. Yet moving the energy index within
a reasonable range ($0.7\approxlt\alpha\approxlt 1.0$) results in the
flux ratio changing from $\sim 0.5$ to $\sim 1$, which does not
explain the ratios observed for the sources in figure 1, some of which
are greater than one and others much smaller than 0.5.

Therefore the dispersion observed in the colour--colour plot seems to
require a more complex spectrum. To solve the problem of high flux
ratios we added a third component to the spectral model just to
increase the soft emission. We modified the previous spectrum with a
blackbody component with temperature $kT=0.1$ keV which is similar to
temperatures fitted to the soft excess emission of some AGN (Kaastra,
Kunieda \& Awaki 1991, Turner et al 1992a, Matsuoka 1994 and references
therein ).  The result is represented in figure 2 as a dashed line
which was produced by assuming $\alpha=0.9$ and by changing the column
density of absorbing gas from Galactic to $10^{24}\rm cm^{-2}$.

By comparing the models above (Figure 2) and the data (Figure 1), we
still see that for an absorbed spectrum, the PSPC hardness ratios are
often much smaller than would be predicted from the 0.5-2 keV to 2-10 keV
flux ratio. The simplest model that can modify this is a partial
covering model, where the absorbing gas only covers a fraction
$f_{cov}$ of the source towards the observer.  This `minimal' spectral
model is then

\begin{equation}
 F(E)= ( A_1 \,{ E^3\over e^{E/kT}-1 } + A_2\, E^{-\alpha})
\end{equation}
$$	 [1 - f_{cov} + f_{cov}\, e^{-\sigma(E)\,N_{H}}]  $$
where $\sigma(E)$ is the absorption cross-section. $A_1$ and
$A_2$ give relative weights to the power law and the black body. 

In order to interpret the data in figure 1, we kept the power law
energy index $\alpha= 0.9$ constant, the blackbody temperature $kT=
0.1$ keV and its relative contribution (50\% of the power law at 1
keV) constant and repeated the simulations for different values of the
column density and the covering factor parameters. The result of the
simulations is presented in figure 1 with solid and dotted lines. Each
solid line corresponds to a different covering factor from 20\%
(bottom) to 100\% (top) and is generated with different values for the
column density that grow from right to left along the curve from
$10^{20} \rm to 3\times 10^{23}\>\rm cm^{-2}$. Each dotted line is drawn
keeping constant the column density whereas the covering factor grows
from bottom to top. The $N_{H}$ value ranges from $10^{21}$ cm$^{-2}$
(first dotted line on the right) to $3\times10^{23}$ cm$^{-2}$ (dotted
line on the left). 
%The AGN with the lowest 0.5-2 keV to 2-10 keV ratio
%(mostly Seyfert 2) have hardness ratios slightly smaller than 1,
%unlike the model predictions. The most likely explanation for this has
%to be found in the contribution of the host galaxy which is expected
%to be rather hard in the {\it Rosat} band (Kim, Fabbiano \& Trinchieri
%1992) and of the same order as the nuclear emission at these soft
%energies.

We also based on this model to correct the hardness ratios for the
effects of Galactic absorption due to its simplicity although as it
will be shown at the end of the section there is an alternative model
able to describe the behaviour of the broad band X--ray colours. In
order to do this correction we defined a model spectrum just as the
one presented above: a power law (with the energy index fixed to 0.9),
a black body (temperature of 0.1 keV and relative contribution of 50\%
also fixed) both partially absorbed by an intrinsic column density
plus an extra absorption representing the effect of the Galaxy. We
folded it through the ROSAT response matrix and repeated the
simulations for a grid of values of the intrinsic column density and
the covering factor and for the values of the Galactic absorption
presented by the sources in the sample. A plot similar to that in
Figure~1 was then obtained for each source (i.e, for each Galactic
column density). The estimated values of the spectral parameters
involved (partial covering factor and intrinsic column density) were
obtained just doing a bilinear interpolation. With these ``true''
parameters we folded again the spectrum for each source so as to get
the corrected Hardness ratio.

Some of the sources (appearing as arrows in the plot) were not
corrected since they fall off this simple model with the 
parameters fixed above requiring a model with slightly different values of the
power law energy index and/or black body temperature and relative
contribution. Thus these sources are shown as upper limits since
the Galactic absorption tends to soften the hardness ratio and to lower
the flux ratio.

Concerning the AGN type (type 1 and type 2) it appears that each class
requires different parameters. While the Seyfert 1 and QSO show a
greater dispersion in the parameters, Seyfert 2 tend to accumulate
around a covering factor of the order of 100\% and/or high column
densities (with the exception of NGC~1068) in agreement with
observations of individual objects (Awaki 1992).  For the majority of
the type 1 AGNs in our sample we infer an absorbing column $\approx
10^{22}\, {\rm cm}^{-2}$ and a covering fraction $f_{cov}$ between
80\% and 100\%. The range over which this last parameter varies is
particularly sensitive to the assumed blackbody temperature (i.e., at
lower blackbody temperatures the unabsorbed spectrum is softer and
therefore more coverage is required), but in any case there is always
an important fraction of the type 1 AGN which require partial
coverage.

If we restrict ourselves to the LMA sources (the LMA sample is flux
limited at 5 keV), the average covering fraction for the type 1 AGN is
%$0.975^{+0.005}_{-0.006}$ 
$0.818^{+0.031}_{-0.036}$ while for the type 2 AGN the average
covering fraction is
%$0.9993^{+0.0017}_{-0.0016}$
$0.9985\pm 0.0007$. It has to be said,
however, that although this subsample is selected in hard X-rays and
therefore there should not be strong biases towards low covering
factors, we only use those AGN for which there is information in the
WGACAT point source catalogue. This might introduce a slight bias
towards low covering factors, but we do not expect it to be too
strong, except for the objects with the largest absorbing columns
(essentially the Seyfert 2s).  Indeed, for soft X-ray selected samples
of AGN, the average covering factor is expected to be larger.

\begin{figure*}
  \vbox to 0cm{\vfil}
  \label{fig3}
\epsfverbosetrue
\epsfysize= 300pt
\epsffile{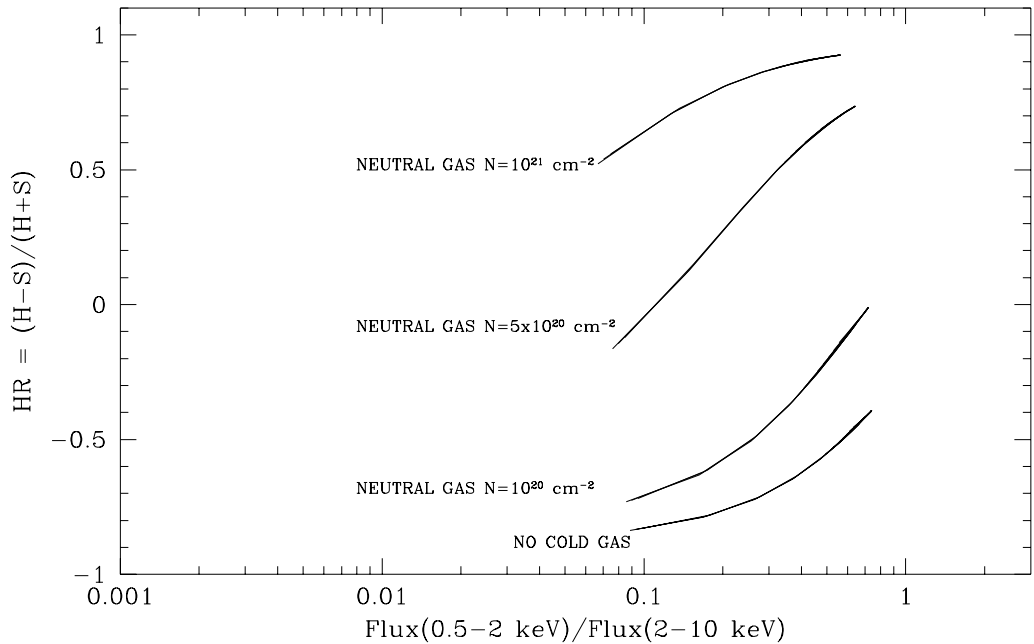}
\caption{ Simulated hardness ratios versus flux ratios for a warm
absorber model (see text). The three uppermost curves correspond to
neutral gas added to the warm component.}
\end{figure*}

We tried to find a similar description of the data in terms of a warm
absorber model (Turner et al 1991, Netzer 1993 and references therein,
Reynolds and Fabian 1995). We used the photoionization code XSTAR
(Kallman and McCray 1982, Kallman and Krolik 1993) to reproduce the
conditions of a thin shell of gas ionized by a primary continuum of
energy index 0.9 and luminosity $L = 4\times 10^{43} \rm erg/s$. The
gas density was fixed to $n=10^9 \rm cm^{-3}$ and XSTAR was run for
different ionization parameters ranging from $\xi=20$ to $\xi=40$
where $\xi\equiv L/(nR^2)$. We selected the XSTAR output to give the
fractional abundance of the ions in the gas (relative to the total
hydrogen abundance), their K-edge energy and their photoionization
cross sections. With these values we constructed a spectrum model for
each gas state (defined by a ionization parameter and a column density
of neutral hydrogen $N_{H}$). This spectrum was folded through the
ROSAT PSPC response matrix in XSPEC to calculate the hardness ratios
just as it was done with the partial covering model. The spectrum
consisted of a power law of energy index $\alpha=0.9$ and an
absorption edge for every ion with significant depth at the threshold
($\tau\geq 0.01$) for a given $N_{H}$. The values of the $N_{H}$
ranged from $10^{19}\, \rm to 10^{22}\, \rm cm^{-2}$). The resulting
hardness ratio versus flux ratio relation is showed in figure 3 as a
solid line.

As it is shown in this figure the variety of X--ray colours cannot be
accounted for only with this model of ionized gas. Thus, in order to obtain
``harder'' colours we added a  neutral gas component to the spectra
defined above. The results for three equivalent neutral hydrogen column
densities in addition to the warm absorber component are presented in figure 3.

We show that in order to reproduce the variety of broad band X--ray
colours observed in the sample a new component of neutral gas must be
added to the partly ionized gas.

\section{LUMINOSITY FUNCTIONS}

We derived the local luminosity function in the hard band (2-10 keV)
so as to compare it with the luminosity function coming from the soft
band (Boyle et al 1994, Maccacaro et al 1991). This can be used to
test whether the simple model presented in this paper predicts a
correct link between soft and hard X-ray energies.

\begin{figure*}
  \vbox to 0cm{\vfil}
  \label{fig4}
\epsfverbosetrue
\epsfysize=300pt
\epsffile{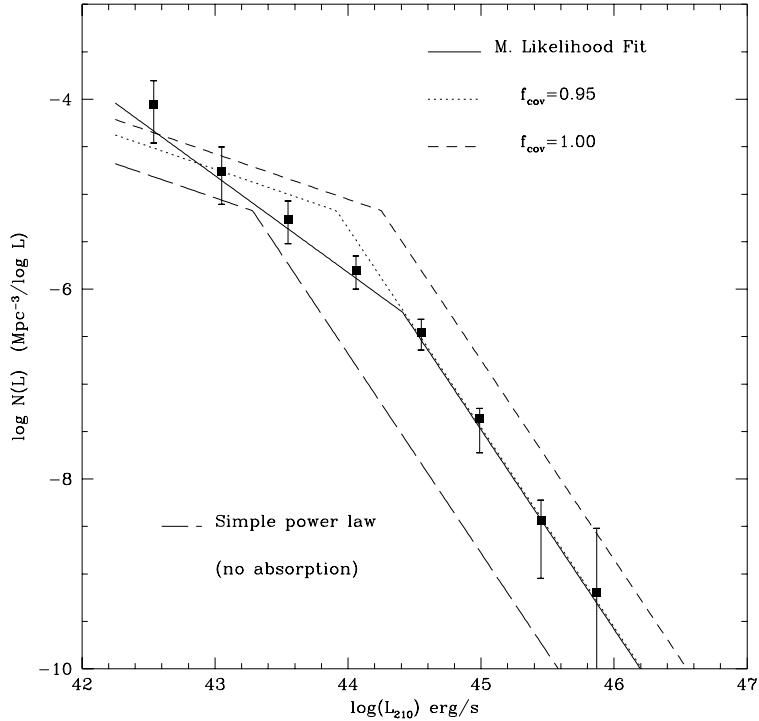}
  \caption{2--10~keV local ($z<0.2$) Luminosity Function. The solid
line represents the maximum likelihood fit to the hard band
sample. The Maccacaro et al. (1991) soft luminosity function was moved
into this hard band assuming a power law energy index $\alpha=0.9$, a
blackbody temperature and contribution of $kT=0.1$~keV and 50\% at 1
keV respectively, $N_{H}=10^{22}\,\rm cm^{-2}$ and two different
covering factors: 0.95(long-dashed line) and 1.0 (dot-dashed
line). The points with the error bars reproduce the nonparametric fit
to the sample with the $1/V_a$ method.}
\end{figure*}

The local luminosity function in the 2-10 keV band is derived from the
whole Grossan (1992) LMA sample (irrespective on whether or not {\it
Rosat} observations exist) restricted to those sources with redshifts
lower than 0.2 and turning to the technique of maximum likelihood
analysis used by Marshall et al. (1983). In his analysis of the
luminosity function for the whole sample, Grossan showed that a
single power law fit was only good as an approximation but that it was
unacceptable for the full range of luminosities since the luminosity
function steepens at high luminosities. Therefore we decided to use a
broken power law form to fit the local hard band luminosity:
\begin{equation}
\Phi(L_{44})= \cases{K_1\, L_{44}^{-\gamma_1} &if $L_{44}\le L_{br}$\cr
\noalign{\medskip}
		       K_2\, L_{44}^{-\gamma_2} &if $L_{44}> L_{br}$\cr}
\end{equation}
where $K_1$ is the normalization of the luminosity function
and $\gamma_1$ and $\gamma_2$ are the faint and the bright end slopes
respectively. The constant $K_2$ is related to the normalization value
through the `break' luminosity, $L_{br}$:
\begin{equation}
K_2 = K_1\, L_{br}^{\gamma_2 - \gamma1}
\end{equation}
$L_{44}$ is the 2-10 keV X-ray luminosity function expressed
in units of $10^{44}$ erg/s. 

In spite of the fact that this sample is reduced to almost local
sources ($z\le 0.2$) the luminosity of the sources was de-evolved to
$z=0$ using a power law evolution form (pure luminosity evolution):
\begin{equation}
L_{44}=L_{44}(z)\,(1+z)^{-\beta}
\end{equation}
with $\beta=2.6$ (Maccacaro et al 1991).

The best-fit values from the maximum likelihood analysis are
$\gamma_1=2.11^{+0.08}_{- 0.09}$, $\gamma_2=3.27^{+0.41}_{-0.25}$, 
$\log(L_{br})= 44.51^{+0.11}_{-0.10}$ and $K_1 = 6.50 \times 10^{-7} 
\,\rm Mpc^{-3}\,(10^{44} \,\rm erg/s)^{-1}$. We tested the goodness of fit
 of the model fit to the data applying a KS statistic a high level of
acceptability.

The comparison between the two luminosity functions is shown in
Figure~4, which displays the binned 2-10 keV luminosity function
together with the maximum likelihood fit and some models.  As usual,
we assume a broken power law for the soft X-ray luminosity function,
and the specific parameters are taken from the {\it Einstein
Observatory} Extended Medium Sensitivity Survey (Maccacaro et al
1991). We prefer this survey to the Boyle et al (1994) one as a
comparison with our 2-10 keV sample, because the Boyle et al (1994)
sample contains virtually no objects at redshifts $z<0.4$ and
therefore the local luminosity function derived in that paper is much
affected by the specific evolution models.  The soft-to-hard band
conversion has been done with a set of `averaged' values for the
parameters of the partial covering spectral model: $\alpha=0.9$,
$kT=0.1$ keV, a blackbody contribution $\sim 50\%$ at 1 keV, $N_{H}=
10^{22}\,\rm cm^{_2}$ and different covering factors. Again, we must
stress here that we use this particular model due to its simplicity.
The same conclusions should be obtained through the neutral plus
ionized gas model since both models are able to reproduce the dispersion
observed in the X--ray colours versus flux ratios relation. The points with
the error bars in the plot correspond to a nonparametric
representation of the 2 - 10 keV luminosity function obtained with the
$1/V_a$ method developed by Avni and Bachall (1980).

\begin{figure*}
  \vbox to 0cm{\vfil}
  \label{fig5}
\epsfverbosetrue
\epsfysize=300pt
\epsffile{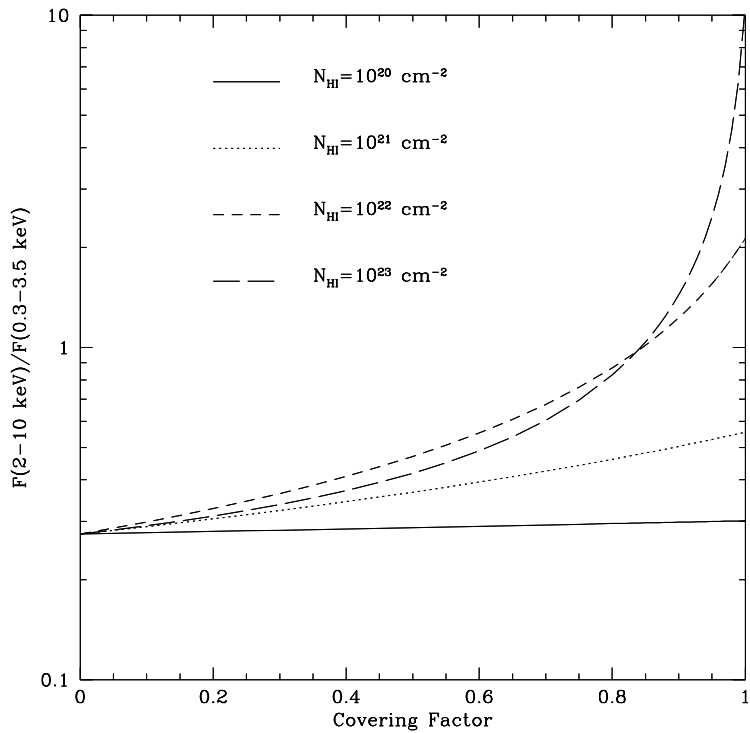}
\caption{Soft (0.3-3.5 keV) to hard (2-10 keV) X--ray
flux ratio versus covering factor. A spectral model with a power law
($\alpha=0.9$), a blackbody emission ($kT=0.1$ keV, 50\% of the power
law at 1 keV) both partially absorbed by different column densities is
assumed to draw the lines (solid line for $N_{H}=10^{20} \,\rm
cm^{-2}$, dotted line for $N_{H}=10^{21} \,\rm cm^{-2}$, short-dashed
line for $N_{H}=10^{22} \,\rm cm^{-2}$ and long-dashed line for $N_{H}=10^{23} \,\rm cm^{-2}$.}
\end{figure*}

As it can be seen from Figure~4 a perfect agreement is rather
difficult to obtain due to the large number of variable parameters
involved in the description of the sources spectrum, specially at and
below the break.  Once we fixed some parameters as blackbody
temperature and its fractional contribution, power law energy index
and absorption column density, it can be seen that very good agreement
between both functions at luminosities $>10^{44}\, {\rm erg}\, {\rm
s}^{-1}$ is reached for a covering factor $\sim 0.95$.  The fit is
poorer for luminosities lower than $L_{2-10}\sim 10^{44}\, {\rm erg}\,
{\rm s}^{-1}$ because it is not possible to introduce the dispersion
observed in the values of some parameters as the column density, the
covering factor and the blackbody contribution, particularly at low
luminosities where we have very few objects. For example, it is likely
that either the covering fraction and/or the absorbing column vary
with luminosity (i.e., low luminosity objects are expected to be more
absorbed), although we cannot detect such effect at a significant level
in the sample discussed in Section 2. A simple model with constant
covering factor and absorbing column is already able to produce very
close agreement between the AGN luminosity functions in the soft and
hard bands, but it would require some refinement when the data samples
can be enlarged.

\section{DISCUSSION}

We have studied the hardness ratios for a sample of hard X-ray
selected sources, all of them observed by ROSAT, and derived a source
emission spectrum that would explain their X--ray colours. This
spectrum has three components: a power law with an energy index
$\alpha\sim 0.9$, a blackbody of $kT\sim 0.1$ keV representing about
50\% of the power law at 1~keV and a low-energy absorption by neutral
gas with column density $N_{H}\sim 10^{22}-10^{23}\> \rm cm^{-2}$
that partially covers the source ($f_{cov}\sim 80-100\%$). Type 2 AGN
appear to be fully covered with large column densities while type 1
AGN have an average covering factor $\sim 0.82$ and absorbing column
density $\approx 10^{22}\, {\rm cm}^{-2}$.  Qualitatively this model
can account for the scatter observed in the hardness ratios as well as
for the soft excess found in some sources.

A complementary result of our analysis is that starting
from a single population of AGN with a distribution of covering
factors, it is possible to describe the two distinct populations
proposed by Franceschini et al (1993). Their soft X--ray class of
active galaxies, showing steep power law spectra and being easily
detected by soft X-ray band missions would correspond to those sources
that in our model had lower covering factors and eventually exhibit
soft excess emission. The hard population would be composed of those
sources strongly self-absorbed by high covering factors. The selection
of objects in the 2-10 keV band does not particularly favour high
values of the covering factor, as it is demonstrated by the presence
of a large fraction of partially covered type 1 AGN in the sample used
in Section 2. However, for a soft X-ray selected sample, the average
covering factor will certainly be smaller. Our model is more alike the
one used by Comastri et al (1995) where a single AGN population is
used to reproduce the spectrum of the X-ray background.

As far as the mismatch between the number counts in different X-ray
bands is concerned, a typical spectrum with a 50\% blackbody
contribution at 1 keV, a power law energy index $\alpha\sim 0.9$, an
absorbing column density between $N_{H}\sim 10^{22}$ and $N_{H}\sim
10^{23}$ and a covering factor $f\approxgt 85\%$ would result in a
flux ratio $F(2-10\,\rm keV)/F(0.3-3.5\, \rm keV)\sim 2$ (see Figure
5) which is the required value to solve the discrepancy.  Thus, the
model presented here with the average parameters obtained in Section 2
also brings the soft and hard source counts into agreement.

We also tried to describe the dispersion on the X--ray colours through
a warm absorber model and the conclusion we can extract from this
analysis is that the presence of gas in two phases (neutral and partly
ionized) is required to reproduce the X--ray colours of most of the
sampled AGN.

Finally, we can derive an interesting conclusion from this analysis:
the gas responsible for the absorption of the X--ray primary spectrum
emitted by the active nucleus must have structure. This structure
could be due to holes or to the coexistence of gas in two phases
(neutral and ionized).

This research has made use of data obtained through the High Energy
Astrophysics Science Archive Research Center Online Service, provided
by the NASA-Goddard Space Flight Center. We acknowledge J.P.D.  Mittaz
and F.J. Carrera at Mullard Space Science Laboratory for help and Ruth
Carballo for careful reading. We also thank the referee for helpful
comments and A. Fabian for interesting suggestions. Partial financial
support for this research was provided by the Spanish DGICYT under
project PB92-0501. MTC was supported by a fellowship from the
Universidad de Cantabria.

\clearpage
%\section{FIGURE CAPTIONS}

%\begin{itemize}
%\item{}Figure~1: 

%\item{}Figure~2: 

%\item{}Figure~3: 

%\item{}Figure~4: 

%\item{}Figure~5: 

%\end{itemize}
%\pagebreak

% Table 1 

\begin{table}
\centering
 \begin{minipage}{22cm}
  \caption{The sample}
  \label{LMA-sample}
  \begin{tabular}{ | l c c c c c c | }\hline
 Name & redshift & HR$^a$ & $F_x(0.5-2)^b$ &
 $F_x(2-10)^c$ & Type$^d$ &  Ref$^e$ \\ \hline
3C273         & 0.1580 & 0.089$\pm$0.010 & 5.7065$\pm$0.047 & 6.5356$\pm$0.559 & 1 & LMA \\
3C390.3       & 0.0570 & 0.824$\pm$0.020 & 0.7459$\pm$0.029 & 1.8257$\pm$0.152 & 1 & LMA \\
AKN374        & 0.0640 & -0.103$\pm$0.015 & 0.5557$\pm$0.010 & 2.4291$\pm$0.405 & 1 & LMA \\
&&&&&&\\
ES0-141-G55   & 0.0370 & 0.637$\pm$0.012 & 2.0457$\pm$0.032 & 2.6798$\pm$0.251 & 1 & LMA \\
Fairall9      & 0.0450 & 0.172$\pm$0.010 & 2.5278$\pm$0.027 & 3.6052$\pm$0.405 & 1 & LMA \\
H0439-272     & 0.0800 & 0.204$\pm$0.028 & 0.4865$\pm$0.015 & 2.5834$\pm$0.405 & 1 & LMA \\
&&&&&&\\
H1029-140     & 0.0860 & 0.635$\pm$0.007 & 1.9795$\pm$0.020 & 2.0243$\pm$0.347 & 1 & LMA \\
H1318+692     & 0.0680 & 0.062$\pm$0.066 & 0.1298$\pm$0.010 & 1.9857$\pm$0.308 & 1 & LMA \\
H1320+551     & 0.0640 & 0.054$\pm$0.041 & 0.1971$\pm$0.009 & 2.0821$\pm$0.405 & 1 & LMA \\
&&&&&&\\
H1419+480     & 0.0720 & -0.064$\pm$0.053 & 0.2569$\pm$0.016 & 2.5834$\pm$0.347 & 1 & LMA \\
IC4329A       & 0.0160 & 0.986$\pm$0.001 & 2.8368$\pm$0.028 & 6.3813$\pm$0.617 & 1 & LMA \\
MCG-2-58-22   & 0.0470 & 0.367$\pm$0.013 & 3.1853$\pm$0.047 & 3.0846$\pm$0.347 & 1 & LMA \\
&&&&&&\\
MCG-6-30-15   & 0.0080 & 0.650$\pm$0.009 & 1.8857$\pm$0.024 & 4.3185$\pm$0.559 & 1 & LMA \\
MKN279        & 0.0310 & -0.079$\pm$0.008 & 2.3228$\pm$0.024 & 3.0846$\pm$0.212 & 1 & LMA \\
MKN290        & 0.0290 & 0.190$\pm$0.030 & 0.5307$\pm$0.018 & 2.1207$\pm$0.251 & 1 & LMA \\
&&&&&&\\
MKN352        & 0.0150 & 0.877$\pm$0.033 & 0.0997$\pm$0.008 & 2.7954$\pm$0.347 & 1 & LMA \\
MKN376        & 0.0560 & 0.875$\pm$0.113 & 0.0131$\pm$0.003 & 3.6052$\pm$0.405 & 1 & LMA \\
MKN464        & 0.0510 & 0.393$\pm$0.053 & 0.2469$\pm$0.015 & 3.2967$\pm$0.463 & 1 & LMA \\
&&&&&&\\
MKN478        & 0.0790 & -0.642$\pm$0.024 & 0.7603$\pm$0.047 & 1.8701$\pm$0.559 & 1 & LMA \\
MKN506        & 0.0430 & 0.447$\pm$0.053 & 0.3304$\pm$0.020 & 1.8257$\pm$0.193 & 1 & LMA \\
MKN509        & 0.0350 & 0.404$\pm$0.012 & 3.8240$\pm$0.053 & 5.3210$\pm$0.617 & 1 & LMA \\
&&&&&&\\
MKN705        & 0.0280 & 0.344$\pm$0.022 & 1.0015$\pm$0.025 & 1.9279$\pm$0.308 & 1 & LMA \\
MKN876        & 0.1290 & 0.132$\pm$0.022 & 0.4392$\pm$0.011 & 1.9279$\pm$0.251 & 1 & LMA \\
MKN1152       & 0.0520 & -0.018$\pm$0.023 & 0.9415$\pm$0.026 & 3.0846$\pm$0.347 & 1 & LMA \\
&&&&&&\\
MR2251-178    & 0.0680 & 0.141$\pm$0.016 & 2.0315$\pm$0.037 & 2.7954$\pm$0.617 & 1 & LMA \\
NGC985        & 0.0430 & 0.150$\pm$0.019 & 0.8286$\pm$0.018 & 1.8257$\pm$0.308 & 1 & LMA \\
NGC2992       & 0.0070 & 1.000$\pm$0.000 & 0.0853$\pm$0.016 & 4.9740$\pm$0.405 & 2 & LMA \\
&&&&&&\\
NGC3227       & 0.0030 & 0.899$\pm$0.007 & 0.4239$\pm$0.007 & 4.4149$\pm$0.501 & 1 & LMA \\
NGC3783       & 0.0090 & 0.865$\pm$0.013 & 1.3855$\pm$0.040 & 3.2389$\pm$0.463 & 1 & LMA \\
NGC4151       & 0.0030 & 0.206$\pm$0.012 & 0.3234$\pm$0.004 & 4.0100$\pm$0.463 & 1 & LMA \\
&&&&&&\\
NGC4593       & 0.0090 & 0.220$\pm$0.040 & 0.9621$\pm$0.042 & 4.2606$\pm$0.655 & 1 & LMA \\
NGC5033       & 0.0030 & 0.354$\pm$0.044 & 0.1768$\pm$0.009 & 3.3931$\pm$0.868 & 2 & LMA \\
NGC5506       & 0.0070 & 0.957$\pm$0.012 & 0.2373$\pm$0.010 & 1.8701$\pm$0.251 & 2 & LMA \\
&&&&&&\\
NGC5548       & 0.0170 & -0.154$\pm$0.008 & 3.3307$\pm$0.034 & 6.4392$\pm$0.405 & 1 & LMA \\
NGC7172       & 0.0080 & 0.355$\pm$0.273 & 0.0056$\pm$0.002 & 1.8701$\pm$0.405 & 2 & LMA \\
NGC7213       & 0.0060 & 0.129$\pm$0.009 & 3.5311$\pm$0.035 & 3.2967$\pm$0.463 & 1 & LMA \\
&&&&&&\\
NGC7469       & 0.0170 & 0.811$\pm$0.005 & 1.5661$\pm$0.015 & 4.9740$\pm$0.501 & 1 & LMA \\
NGC7582       & 0.0050 & 0.725$\pm$0.061 & 0.0370$\pm$0.003 & 4.7619$\pm$0.405 & 2 & LMA \\
NGC7674       & 0.0290 & 0.467$\pm$0.181 & 0.0146$\pm$0.003 & 2.5448$\pm$0.559 & 2 & LMA \\
&&&&&&\\
PG0052+251    & 0.1540 & 0.478$\pm$0.022 & 0.5366$\pm$0.014 & 4.0100$\pm$0.713 & 1 & LMA \\
PG0804+76     & 0.1000 & 0.265$\pm$0.025 & 0.8677$\pm$0.024 & 2.1207$\pm$0.308 & 1 & LMA \\
PictorA       & 0.0340 & 0.500$\pm$0.335 & 0.0035$\pm$0.001 & 2.3327$\pm$0.308 & 1 & LMA \\
&&&&&&\\ \hline
   \end{tabular}
\end{minipage}
\end{table}

\begin{table*}
 \begin{minipage}{22cm}
  \caption*{\it (continued)}
  \label{OTRA-sample}
  \begin{tabular}{ | l c c c c c c | }\hline
 Name & redshift & HR\footnote{Hardness Ratio not corrected for Galactic absorption} & $F_x(0.5-2)$\footnote{Flux
 in $10^{-11} \ergpcmsqps$ in the (0.5-2) keV band} &
 $F_x(2-10)$\footnote{Flux in $10^{-11} \ergpcmsqps$ in the (2-10)
 keV band} & Type\footnote{Type: 1 for Sy1/QSO and 2 for Sy2}
 &  Ref\footnote{References for 2-10 keV Flux: LMA (Grossan 1992), EXO 
(Turner \& Pounds 1989), \newline
   GINGA (Nandra \& Pounds 1994, Turner et al 1992b, Awaki 1992)} \\ \hline
3C120         & 0.0330 & 0.974$\pm$0.005 & 2.8433$\pm$0.069 & 4.3000$\pm$0.430 & 1 & EXO \\
MKN590        & 0.0270 & 0.145$\pm$0.013 & 3.2585$\pm$0.047 & 2.7000$\pm$0.270 & 1 & EXO \\
NGC1068       & 0.0030 & 0.120$\pm$0.037 & 1.1321$\pm$0.047 & 0.5300$\pm$0.050 & 2 & EXO \\
NGC6814       & 0.0050 & 0.965$\pm$0.005 & 0.2315$\pm$0.005 & 3.0000$\pm$0.300 & 1 & EXO \\
&&&&&&\\
3C111         & 0.0480 & 0.991$\pm$0.009 & 0.6151$\pm$0.049 & 3.1300$\pm$0.190 & 1 & GINGA \\
AKN120        & 0.0330 & 0.925$\pm$0.005 & 1.8001$\pm$0.024 & 3.5800$\pm$0.190 & 1 & GINGA \\
IIZw136       & 0.0630 & 0.205$\pm$0.017 & 0.8046$\pm$0.015 & 0.4800$\pm$0.000 & 1 & GINGA \\
MCG-5-23-16   & 0.0080 & 0.957$\pm$0.011 & 0.3378$\pm$0.014 & 2.2400$\pm$0.180 & 2 & GINGA \\
&&&&&&\\
MKN3          & 0.0137 & 0.928$\pm$0.016 & 0.0797$\pm$0.004 & 0.8800$\pm$0.000 & 2 & GINGA \\
MKN205        & 0.0710 & 0.350$\pm$0.025 & 0.5764$\pm$0.016 & 0.9000$\pm$0.000 & 1 & GINGA \\
MKN335        & 0.0250 & 0.015$\pm$0.009 & 1.6386$\pm$0.016 & 1.2500$\pm$0.150 & 1 & GINGA \\
MKN348        & 0.0150 & 0.813$\pm$0.060 & 0.0094$\pm$0.001 & 1.1000$\pm$0.000 & 2 & GINGA \\
&&&&&&\\
MKN372        & 0.0310 & 0.903$\pm$0.010 & 0.3270$\pm$0.009 & 0.2000$\pm$0.000 & 2 & GINGA \\
MKN841        & 0.0360 & 0.020$\pm$0.021 & 1.5708$\pm$0.038 & 0.7900$\pm$0.160 & 1 & GINGA \\
NGC1808       & 0.0030 & 0.925$\pm$0.020 & 0.0636$\pm$0.004 & 0.3000$\pm$0.000 & 2 & GINGA \\
NGC3516       & 0.0090 & 0.305$\pm$0.007 & 3.4997$\pm$0.025 & 2.0900$\pm$0.190 & 1 & GINGA \\
&&&&&&\\
NGC4051       & 0.0020 & -0.247$\pm$0.026 & 0.3700$\pm$0.013 & 1.9500$\pm$0.170 & 1 & GINGA \\
NGC7314       & 0.0060 & 0.969$\pm$0.004 & 0.2243$\pm$0.004 & 2.8300$\pm$0.180 & 2 & GINGA \\
PG1211+143    & 0.0850 & -0.162$\pm$0.031 & 0.6110$\pm$0.024 & 0.7700$\pm$0.000 & 1 & GINGA \\
PG1307+085    & 0.1550 & -0.048$\pm$0.029 & 0.2487$\pm$0.009 & 0.6400$\pm$0.000 & 1 & GINGA \\
&&&&&&\\
PG1352+183    & 0.1520 & -0.268$\pm$0.025 & 0.2087$\pm$0.007 & 0.1600$\pm$0.000 & 1 & GINGA \\
PG1416-129    & 0.1290 & 0.879$\pm$0.012 & 0.4032$\pm$0.011 & 0.6600$\pm$0.000 & 1 & GINGA \\
PHL1657       & 0.2000 & 0.498$\pm$0.054 & 0.3366$\pm$0.022 & 1.1800$\pm$0.000 & 1 & GINGA \\ \hline
   \end{tabular}
\end{minipage}
\end{table*}

\end{document}